\documentclass[12pt,journal,onecolumn,doublespacing]{IEEEtran}
\usepackage{url}
\usepackage{setspace}
\usepackage{amsmath}
\usepackage{multicol}
\usepackage{amssymb}
\usepackage[utf8]{inputenc}
\usepackage[T1]{fontenc}
\usepackage[font={small}]{caption}
\usepackage{cite}
\usepackage{enumerate}
\usepackage[square, comma, numbers, sort&compress]{natbib}
\usepackage{subfig}
\usepackage{bm}
\usepackage{multirow}
\usepackage{xcolor}
\usepackage{textcomp}

\ifCLASSINFOpdf
\else
\fi
\ifCLASSINFOpdf
   \usepackage[pdftex]{graphicx}
   \usepackage{epstopdf}
\else
\fi

\begin{document}
	
\title{\huge Multi-Antenna Relaying and Reconfigurable Intelligent Surfaces: End-to-End SNR and Achievable Rate}
\author{\normalsize K.~Ntontin,~\IEEEmembership{\normalsize Member,~IEEE},
\normalsize J.~Song, and
\normalsize M.~Di~Renzo,~\IEEEmembership{\normalsize Senior~Member,~IEEE} \vspace{-0.65cm}
\thanks{Received Aug. 31, 2019. K. Ntontin is with Demokritos, Athens, Greece. J. Song and M. Di Renzo are with the Laboratoire des Signaux et Syst\`emes, CNRS, CentraleSup\'elec, Univ Paris Sud, Universit\'e Paris-Saclay, 3 rue Joliot Curie, Plateau de Saclay, 91192, Gif-sur-Yvette, France (e-mail: marco.direnzo@l2s.centralesupelec.fr)} }
%
%
\markboth{Unpublished Report} {K. Ntontin, J. Song, and M. Di Renzo: Multi-Antenna Relaying and Reconfigurable Intelligent Surfaces: End-to-End SNR and Achievable Rate}

\maketitle

\begin{abstract}
In this report, we summarize the end-to-end signal-to-noise ratio and {achievable} rate of half-duplex, full-duplex, amplify-and-forward, and decode-and-forward relay-assisted communication, {as} well as the signal-to-noise ratio and {achievable} rate of the emerging technology known as reconfigurable intelligent surfaces. 
\end{abstract}

\section{System Model}

Let us assume that a single-antenna source wishes to communicate with a single-antenna destination that is located at a distance $d_{SD}$. Due to the possibility of the source-to-destination link being of poor quality due to blockages, such as fixed and moving objects, which can frequently be the case for communication at high-frequency bands, such as millimeter wave (30-100 GHz) and sub-millimeter wave (greater than 100 GHz) bands, a multiple-antenna relay \cite{Mischa} or a multiple-element reconfigurable intelligent surface (RIS) \cite{MDR_Eurasip}, \cite{MDR_Access} can assist the communication between the source and the destination, as depicted in Fig.~\ref{System model}. As far as the relay is concerned, we consider both half-duplex (HD) and full-duplex (FD) duplexing schemes, and decode-and-forward (DF) and amplify-and-forward (FD) relaying protocols. The distance of the source-to-relay/RIS link is denoted by $d_{SR}$ and the distance of the relay/RIS-to-destination link is denoted by $d_{RD}$. In scenarios where the source-to-destination link can frequently be under non-line-of-sight (NLOS) propagation conditions, the relay/RIS should be placed at positions where a line-of-sight (LOS) link is secured for both the source-to-relay/RIS and relay/RIS-to-destination links.

Furthermore, we denote the number of antennas and elements of the relay and the RIS by $N_{R}$ and $N_{RIS}$, respectively. In addition, we denote the path-loss at a distance $d$ by $P_{L}(d)$, and the transmitted symbol from the source by $s$. Finally, we consider additive white Gaussian noise affecting the relay and destination, whose variance is equal to ${N_0} =  - 174 + 10lo{g_{10}}\left( {BW} \right) + NF\;\left( {dBm} \right)$, where $BW$ and $NF$ are the signal bandwidth and noise figure, respectively.

{The content of the rest of this technical report is described as follows: In Section~\ref{end_to_end_SNR}, the end-to-end signal-to-noise ratio (SNR) of relay-assisted and RIS-assisted communication systems is derived, and in Section~\ref{achievable_rate} the corresponding maximum achievable rate is presented}.

\begin{figure}[!t]
	\label{System model}
	\centering
	\includegraphics[width=5in,height=5in]{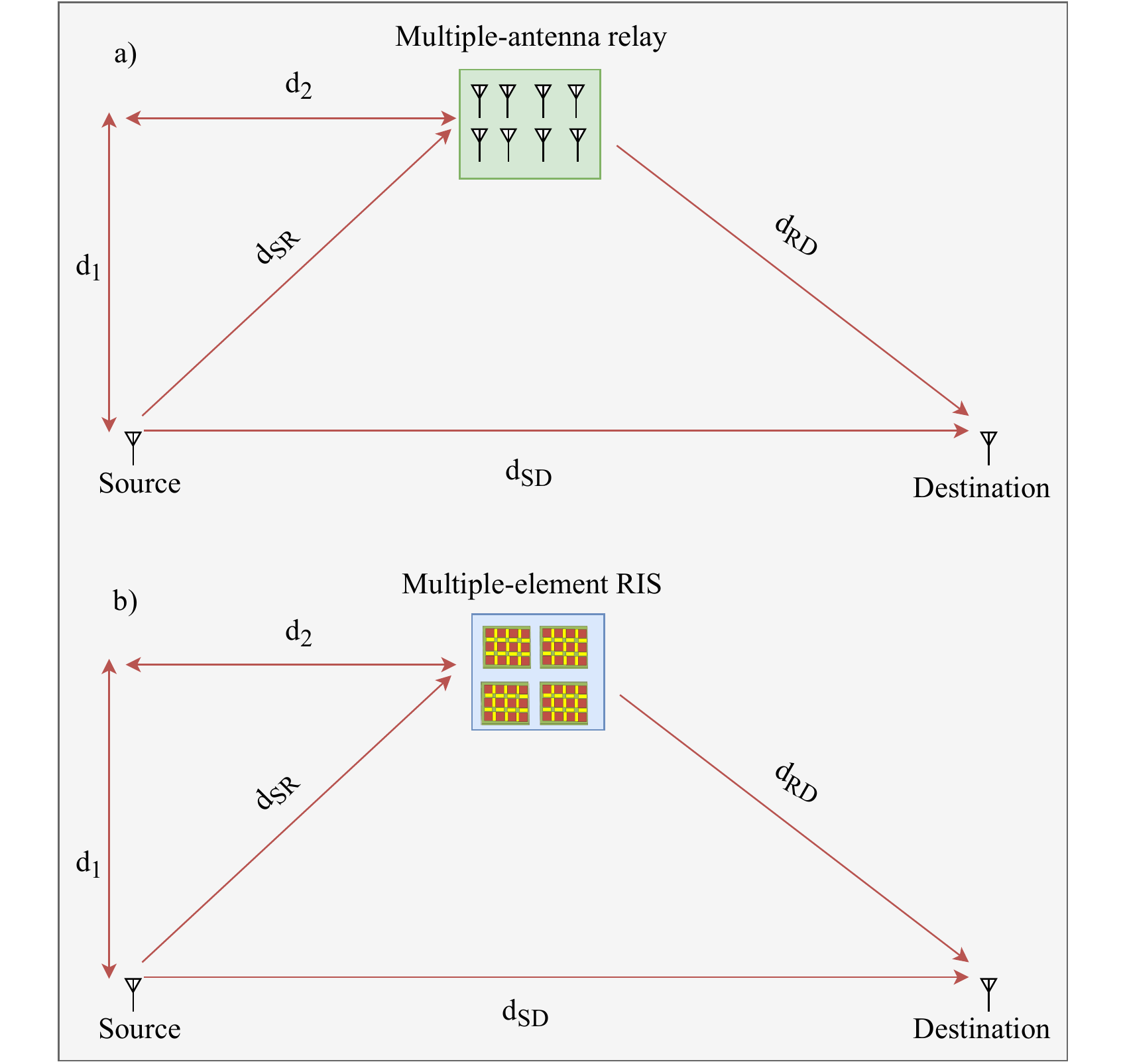}
	\caption{System model: a) Relay-assisted communication and b) RIS-assisted communication}
	\label{System model}
\end{figure}

\section{{End-to-end SNR}}

\label{end_to_end_SNR}

{In this section, we provide the mathematical formulation of the end-to-end SNR for both relay-assisted and RIS-assisted communication}.

\subsection{Relay-Assisted Communication}

In this section, we introduce the mathematical formulation of the end-to-end SNR expressions for the relay-assisted communication. By ${h_{SD}} \in \mathbb{C}$, ${{\bf{h}}_{SR}} \in \mathbb{C}{^{{N_R} \times 1}}$, and ${{\bf{h}}_{RD}} \in \mathbb{C} {^{1 \times {N_R}}}$, we denote the fast-fading complex envelopes of the source-to-destination, source-to-relay, and relay-to-destination links, respectively. We further assume that the relay and the destination have complete knowledge of the instantaneous channel-state information (CSI). This serves as an upper bound of the resulting performance. Finally, with $p_S$ and $p_R$ we denoted the power levels used for the transmission from the source and the relay, respectively. A total power budget equal to $p_{tot}$ is considered, which implies $p_{tot}=p_S+p_R$.

\subsubsection{\textbf{HD Relaying}}

Under HD relaying, the communication between the source and the destination is realized in two time slots. In the first time slot, only the source transmits and its symbol is received by the relay, which performs maximal ratio combining (MRC) \cite{goldsmith2005wireless}, and by the destination. In the second time slot, only the relay transmits by using the maximal ratio transmission (MRT) \cite{goldsmith2005wireless} principle and the signal received at the destination is combined, by leveraging the MRC principle, with the signal received in the first time slot.

\paragraph{DF Relaying} \mbox{}\\ \textbf{First time slot} -- In the first time slot, the signal received at the destination, which we denote by $y_{{D_{DF}}}^{\left( {HD} \right)\left( 1 \right)}$, is given by
\begin{small}
\begin{equation}
\label{1st_slot_DF_destination}
y_{{D_{DF}}}^{\left( {HD} \right)\left( 1 \right)} = \sqrt {p_S{P_L}\left( {{d_{SD}}} \right)} {h_{SD}}s + {n_D^{\left( 1 \right)}}.
\end{equation}
\end{small}
Hence, the SNR at the destination in the first time slot, which we denote by $\gamma_{{D_{DF}}}^{\left( {HD} \right)\left( 1 \right)}$, is given by
\begin{small}
\begin{equation}
\label{1st_slot_SNR_destination}
\gamma _{{D_{DF}}}^{\left( {HD} \right)\left( 1 \right)} = \frac{{{p_S}{P_L}\left( {{d_{SD}}} \right){{\left| {{h_{SD}}} \right|}^2}}}{{{N_0}}}.
\end{equation}
\end{small}
The signal received at the relay, which we denote by $y_{{R_{DF}}}^{\left( {HD} \right)}$, is given by
\begin{small}
\begin{equation}
\label{1st_slot_received_signal_relay}
y_{{R_{DF}}}^{\left( {HD} \right)} = \sqrt {p_S{P_L}\left( {{d_{SR}}} \right)} {{\bf{w}}_{{R_{MRC}}}^{\left(HD\right)}}{{\bf{h}}_{SR}}s + {{\bf{w}}_{{R_{MRC}}}^{\left(HD\right)}}{{\bf{n}}_R},
\end{equation}
\end{small}
where
\begin{small}
\begin{equation}
{{\bf{w}}_{{R_{MRC}}}^{\left(HD\right)}} = \frac{{{\bf{h}}_{SR}^H}}{{\left\| {{{\bf{h}}_{SR}}} \right\|}}
\end{equation}
\end{small}
is the MRC combining vector at the relay. Consequently, the SNR at the relay, which we denote by $\gamma_{{R_{DF}}}^{\left( {HD} \right)}$, is given by
\begin{small}
\begin{equation}
\label{1st_slot_SNR_relay}
\gamma _{{R_{DF}}}^{\left( {HD} \right)} = \frac{{{p_S}{P_L}\left( {{d_{SR}}} \right){{\left\| {{{\bf{h}}_{SR}}} \right\|}^2}}}{{{N_0}}}.
\end{equation}
\end{small}

\mbox{}\\ \textbf{Second time slot} -- Let us denote the remodulated symbol at the relay after decoding by ${\bar s}$. According to the MRT principle, the signal received at the destination in the second time slot, which is denoted by $y_{{D_{DF}}}^{\left( {HD} \right)\left( 2 \right)}$, is given by
\begin{small}
\begin{equation}
y_{{D_{DF}}}^{\left( {HD} \right)\left( 2 \right)} = \sqrt {{p_R}{P_L}\left( {{d_{RD}}} \right)} {{\bf{h}}_{RD}}{{\bf{w}}_{{R_{MRT}}}^{\left(HD\right)}}\bar s + n_D^{\left( 2 \right)},
\end{equation}
\end{small}
where
\begin{small}
\begin{equation}
{{\bf{w}}_{{R_{MRT}}}^{\left(HD\right)}} = \frac{{{\bf{h}}_{RD}^H}}{{\left\| {{{\bf{h}}_{RD}}} \right\|}}
\end{equation}
\end{small}
is the MRT combining vector at the relay.
Hence, the SNR at the destination in the second time slot, which is denoted by $\gamma_{{D_{DF}}}^{\left( {HD} \right)\left( 2 \right)}$, is given by
\begin{small}
\begin{equation}
\label{SNR_relay-destination link}
\gamma _{{D_{DF}}}^{\left( {HD} \right)\left( 2 \right)} = \frac{{{p_R}{P_L}\left( {{d_{RD}}} \right){{\left\| {{{\bf{h}}_{RD}}} \right\|}^2}}}{{{N_0}}}.
\end{equation}
\end{small}
By combining $y_{{D_{DF}}}^{\left( {HD} \right)\left( 1 \right)}$ and $y_{{D_{DF}}}^{\left( {HD} \right)\left( 2 \right)}$ at the receiver, the resulting combined signal, which we denote by $y_{{D_{DF}}}^{\left( {HD} \right)}$, is given by
\begin{small}
\begin{equation}
y_{{D_{DF}}}^{\left( {HD} \right)} = \left[ {\begin{array}{*{20}{c}}
	{y_{{D_{DF}}}^{\left( {HD} \right)\left( 1 \right)}}&{y_{{D_{DF}}}^{\left( {HD} \right)\left( 2 \right)}}
	\end{array}} \right]{{\bf{w}}_{{D_{MRC}}}},
\end{equation}
\end{small}
where
\begin{small}
\begin{equation}
{{\bf{w}}_{{D_{MRC}}}} = \frac{{{{\left[ {\begin{array}{*{20}{c}}
					{{h_{SD}}}&{\left\| {{{\bf{h}}_{RD}}} \right\|}
					\end{array}} \right]}^H}}}{{\left\| {\begin{array}{*{20}{c}}
			{h_{SD}}&{\left\| {{{\bf{h}}_{RD}}} \right\|}
			\end{array}} \right\|}}
\end{equation}
\end{small}
is the MRC combining vector at the destination. Consequently, the resulting SNR at the destination, which we denote by $\gamma_{{D_{DF}}}^{\left( {HD} \right)}$, is given by
\begin{small}
\begin{equation}
\gamma _{{D_{DF}}}^{\left( {HD} \right)} = \frac{{{p_S}{P_L}\left( {{d_{SD}}} \right){{\left| {{h_{SD}}} \right|}^2}}}{{{N_0}}} + \frac{{{p_R}{P_L}\left( {{d_{RD}}} \right){{\left\| {{{\bf{h}}_{RD}}} \right\|}^2}}}{{{N_0}}} = \gamma _{{D_{DF}}}^{\left( {HD} \right)\left( 1 \right)} + \gamma _{{D_{DF}}}^{\left( {HD} \right)\left( 2 \right)}.
\end{equation}
\end{small}

\paragraph{AF Relaying} \mbox{}\\ \textbf{First time slot} -- In the first time slot, the signals received at the destination and the relay, which we denote by $y_{{D_{AF}}}^{\left( {HD} \right)\left( 1 \right)}$ and $y_{{R_{AF}}}^{\left( {HD} \right)}$, are given by \eqref{1st_slot_DF_destination} and \eqref{1st_slot_received_signal_relay}, respectively. Consequently, the corresponding SNR formulas, which we denote by $\gamma_{{D_{AF}}}^{\left( {HD} \right)\left( 1 \right)}$ and $\gamma_{{R_{AF}}}^{\left( {HD} \right)}$, are given by \eqref{1st_slot_SNR_destination} and \eqref{1st_slot_SNR_relay}, respectively.

\mbox{}\\ \textbf{Second time slot} -- By assuming a variable-gain relay, in the second time slot the signal received at the destination, which we denote by $y_{{D_{AF}}}^{\left( {HD} \right)\left( 2 \right)}$, is given by
\begin{small}
\begin{equation}
y_{{D_{AF}}}^{\left( {HD} \right)\left( 2 \right)} = {G_R^{\left(HD\right)}}y_{{R_{AF}}}^{\left( {HD} \right)}\sqrt {{p_R}{P_L}\left( {{d_{RD}}} \right)} {{\bf{h}}_{RD}}{{\bf{w}}_{{R_{MRT}}}^{\left(HD\right)}} + n_D^{\left( 2 \right)},
\end{equation}
\end{small}
where
\begin{small}
\begin{equation}
{G_R^{\left(HD\right)}} = \sqrt {\frac{1}{{{p_S}{P_L}\left( {{d_{SR}}} \right){{\left\| {{{\bf{h}}_{SR}}} \right\|}^2} + {N_0}}}}
\end{equation}
\end{small}
is the gain of the relay. As a result, the resulting SNR at the destination, which we denote by $\gamma _{{D_{AF}}}^{\left( {HD} \right)(2)}$, is given by
\begin{small}
\begin{equation}
\gamma _{{D_{AF}}}^{\left( {HD} \right)(2)} = \frac{{\gamma _{{R_{AF}}}^{\left( {HD} \right)}\gamma _{{D_{AF}}}^{'\left( {HD} \right)\left( 2 \right)}}}{{\gamma _{{R_{AF}}}^{\left( {HD} \right)} + \gamma _{{D_{AF}}}^{'\left( {HD} \right)\left( 2 \right)} + 1}},
\end{equation}
\end{small}
where ${\gamma _{{D_{AF}}}^{'\left( {HD} \right)\left( 2 \right)}}=\gamma _{{D_{DF}}}^{\left( {HD} \right)\left( 2 \right)}$. After MRC combining at the destination of $y_{{D_{AF}}}^{\left( {HD} \right)\left( 1 \right)}$ and $y_{{D_{AF}}}^{\left( {HD} \right)\left( 2 \right)}$, the resulting SNR, which we denote by $\gamma_{{D_{AF}}}^{\left( {HD} \right)}$, is given by
\begin{small}
	\begin{equation}
	\gamma_{{D_{AF}}}^{\left( {HD} \right)}=\gamma_{{D_{AF}}}^{\left( {HD} \right)\left( 1 \right)}+\gamma_{{D_{AF}}}^{\left( {HD} \right)\left( 2 \right)}.
	\end{equation}
\end{small}

\subsubsection{\textbf{FD Relaying}}
Under FD relaying, the signal transmission is realized in two time slots, as in the HD relaying principle. However, in contrast to the HD case, both the source and the relay are allowed to transmit concurrently on the same physical resource, i.e., a time slot \cite{Riihonen_HD_FD_2011}, \cite{Survey_FD_Relaying}. Due to this protocol, by assuming that in time slot $k$ the source transmits the symbol $s_k$, the signal received at the relay is affected by the loop-back self-interference due to the concurrent transmission from the relay of the signal that corresponds to $s_{k-1}$, which is the symbol transmitted by the source in time slot $k-1$. Furthermore, the signal received by the destination from the relay in time slot $k+1$, which corresponds to the symbol $s_k$, is affected by the interference resulting from the concurrent transmission of the symbol $s_{k+1}$ by the source. Hence, after $M$ signaling transmissions the resulting symbol rate under the FD protocol is equal to $\frac{M}{{M + 1}} \to 1$, as $M \to \infty $, which is the same as the symbol rate of a single-input single output (SISO) transmission. Such a rate advantage of FD relaying compared to its HD counterpart comes at the cost of an increased interference, as explained.

If $N_R\ge2$ antennas are available at the relay, a way to perform FD operation is to allocate half of the antennas for reception and the rest half for transmission. Another way is to simultaneously use the $N_R$ antennas for both transmission and reception by using circuits that are called circulators. In this work, we consider the former approach. Accordingly, by ${{\bf{\tilde h}}_{SR}} \in \mathbb{C}{^{{\frac{N_R}{2}} \times 1}}$ and ${{\bf{\tilde h}}_{RD}} \in \mathbb{C}{^{1 \times {\frac{N_R}{2}}}}$ we denote the fast-fading complex envelopes of the source-to-relay and relay-to-destination links, respectively.

As far as the loop-back self-interference is concerned, it consists of two parts: (i) the direct-path loop-back self-interference propagating directly from the transmit to the receive chain. It can either comprise the LOS signal propagating directly from the transmit to the receive antennas in a separate antenna structure or the signal reaching the receive chain due to the circulators' leakage (due to antenna mismatching, for instance) in a shared antenna structure. The power impact of this component can be much larger than the power of the received signal, which renders the realization of FD-relaying challenging. Hence, it is essential that such a component is substantially suppressed so that it is comparable or smaller than the noise level. This can be realized by the use of highly directional antennas or significant isolation between the transmit and receive antennas, as well as analog and digital cancellation techniques \cite{Survey_FD_Relaying}; and (ii) a multipath part due to scattering and reflections from objects in the radio path. We assume that this fast-fading coefficient for each transmit-receive antenna pair of the relay is described by a zero-mean complex Gaussian random variable with variance that depends on the distance-based path-loss and the shadowing effects that the scattering components experience before reaching the receive antennas.

Based on the above, a widespread approach for modeling the loop-back self-interference consists of assuming that the residual loop-back self-interference, which is the remaining interference after all cancellation stages are implemented, is as a complex Rician random variable \cite{Experiment_Characterization_Full_Duplex}. The strong LOS component of such a process represents the residual level of the direct-path loop-back self-interference. By ${{{\bf{\tilde h}}}_{LI}}\in \mathbb{C}{^{{\frac{N_R}{2}} \times 1}}$, we denote the complex envelope representing such a process. Furthermore, we assume that the relay perfectly knows ${{{\bf{\tilde h}}}_{LI}}$ through, for example, an estimation phase prior to data transmission.

\paragraph{DF Relaying} \mbox{}\\ \textbf{First time slot} -- In the first time slot, the signal received at the relay, which we denote by $y_{{R_{DF}}}^{\left( {FD} \right)}$, is given by
\begin{small}
	\begin{equation}
	\label{1st_slot_received_signal_relay_FD}
	y_{{R_{DF}}}^{\left( {FD} \right)} = \sqrt {p_S{P_L}\left( {{d_{SR}}} \right)} {{\bf{w}}_{{R_{MRC}}}^{\left(FD\right)}}{{\bf{\tilde h}}_{SR}}s +\sqrt {{p_R}} {{\bf{w}}_{{R_{MRC}}}^{\left(FD\right)}}{{{\bf{\tilde h}}}_{LI}}\tilde s' + {{\bf{w}}_{{R_{MRC}}}^{\left(FD\right)}}{{\bf{n}}_R},
	\end{equation}
\end{small}
where $\tilde s'$ is the remodulated symbol, which was conveyed by the source in the previous time slot, after decoding and
\begin{small}
	\begin{equation}
	{{\bf{w}}_{{R_{MRC}}}^{\left(FD\right)}} = \frac{{{\bf{\tilde h}}_{SR}^H}}{{\left\| {{{\bf{\tilde h}}_{SR}}} \right\|}}
	\end{equation}
\end{small}
is the MRC combining vector at the relay.
Consequently, the signal-to-interference-plus-noise ratio (SINR) at the relay, which we denote by $\gamma _{{R_{DF}}}^{\left( {FD} \right)}$, is given by
\begin{small}
	\begin{equation}
	\label{1st_slot_SNR_relay_FD}
	\gamma _{{R_{DF}}}^{\left( {FD} \right)} = \frac{{{p_S}{P_L}\left( {{d_{SR}}} \right){{\left\| {{{\bf{\tilde h}}_{SR}}} \right\|}^2}}}{{{{p_R}{\left| {{{\bf{w}}_{{R_{MRC}}}^{\left(FD\right)}}{{{\bf{\tilde h}}}_{LI}}} \right|^2}+N_0}}}.
	\end{equation}
\end{small}

\mbox{}\\ \textbf{Second time slot} --  In the second time slot, the signal received at the destination that corresponds to the decoded and remodulated symbol $\tilde s$, which is denoted by $y_{{D_{DF}}}^{\left( {FD} \right)}$, is given by
\begin{small}
	\begin{equation}
	y_{{D_{DF}}}^{\left( {FD} \right)} = \sqrt {{p_R}{P_L}\left( {{d_{RD}}} \right)} {{\bf{\tilde h}}_{RD}}{{\bf{w}}_{{R_{MRT}}}^{\left(FD\right)}}\bar s + \sqrt {{p_S}{P_L}\left( {{d_{SD}}} \right)} {h_{SD}}s'' + {n_D},
	\end{equation}
\end{small}
where $s''$ is the symbols conveyed by the source in the current time slot and
\begin{small}
	\begin{equation}
	{{\bf{w}}_{{R_{MRT}}}^{\left(FD\right)}} = \frac{{{\bf{\tilde h}}_{RD}^H}}{{\left\| {{{\bf{\tilde h}}_{RD}}} \right\|}}
	\end{equation}
\end{small}
is the MRT combining vector at the relay.
Hence, the resulting SINR at the destination, which we denote by $\gamma_{{D_{DF}}}^{\left( {FD} \right)}$, is given by
\begin{small}
\begin{equation}
\label{SINR_relay_destination_link_AF}
\gamma_{{D_{DF}}}^{\left( {FD} \right)}=\frac{{{p_R}{P_L}\left( {{d_{RD}}} \right){{\left\| {{{\bf{\tilde h}}_{RD}}} \right\|}^2}}}{{{{{p_S}{P_L}\left( {{d_{SD}}} \right){{\left| {{h_{SD}}} \right|}^2}}+N_0}}}.
\end{equation}
\end{small}

\paragraph{AF Relaying} \mbox{}\\ \textbf{First time slot} -- In the first time slot, the signal received at the relay, which we denote by $y_{{R_{AF}}}^{\left( {FD} \right)}$, is given by
\begin{small}
\begin{equation}	
y_{{R_{AF}}}^{\left( {FD} \right)} = \sqrt {p_S{P_L}\left( {{d_{SR}}} \right)} {{\bf{w}}_{{R_{MRC}}}^{\left(FD\right)}}{{\bf{\tilde h}}_{SR}}s +\sqrt {{p_R}} {{\bf{w}}_{{R_{MRC}}}^{\left(FD\right)}}{{{\bf{\tilde h}}}_{LI}} G_R^{\left(FD\right)}y_{{R_{AF}}}^{'\left( {FD} \right)} + {{\bf{w}}_{{R_{MRC}}}^{\left(FD\right)}}{{\bf{n}}_R},
\end{equation}
\end{small}
where $y_{{R_{AF}}}^{'\left( {FD} \right)}$ is the signal received at the relay in the previous time slot and
\begin{small}
\begin{equation}
{G_R^{\left(FD\right)}} = \sqrt {\frac{1}{{{p_S}{P_L}\left( {{d_{SR}}} \right){{\left\| {{{\bf{\tilde h}}_{SR}}} \right\|}^2} + {p_R}{\left| {{{\bf{w}}_{{R_{MRC}}}^{\left(FD\right)}}{{{\bf{\tilde h}}}_{LI}}} \right|^2}+ {N_0}}}}
\end{equation}
\end{small}
is the gain of the variable-gain relay. Consequently, the resulting SINR at the relay, which we denote by $\gamma_{{R_{AF}}}^{\left( {FD} \right)}$, is given by
\begin{small}
	\begin{equation}
\gamma _{{R_{AF}}}^{\left( {FD} \right)} = \frac{{{p_S}{P_L}\left( {{d_{SR}}} \right){{\left\| {{{\bf{\tilde h}}_{SR}}} \right\|}^2}}}{{{{p_R}{\left| {{{\bf{w}}_{{R_{MRC}}}^{\left(FD\right)}}{{{\bf{\tilde h}}}_{LI}}} \right|^2}+N_0}}}.
\end{equation}
\end{small}

\mbox{}\\ \textbf{Second time slot} -- In the second time slot, the signal received at the destination, which we denote by $y_{{D_{AF}}}^{\left( {FD} \right)}$, is given by
\begin{small}
	\begin{equation}
	y_{{D_{AF}}}^{\left( {FD} \right)} = {G_R}y_{{R_{AF}}}^{\left( {FD} \right)}\sqrt {{p_R}{P_L}\left( {{d_{RD}}} \right)} {{\bf{\tilde h}}_{RD}}{{\bf{w}}_{{R_{MRT}}}^{\left(FD\right)}} +\sqrt {{p_S}{P_L}\left( {{d_{SD}}} \right)} {h_{SD}}s''+ n_D.
\end{equation}
\end{small}
After some algebraic manipulations, the SINR at the destination, which we denote by $\gamma_{{D_{AF}}}^{\left( {FD} \right)}$, is given by
\begin{small}
	\begin{equation}
	\gamma _{{D_{AF}}}^{\left( {FD} \right)} = \frac{{\gamma _{{R_{AF}}}^{\left( {FD} \right)}\gamma _{{D_{AF}}}^{\left( {FD} \right)\left( 2 \right)}}}{{\gamma _{{R_{AF}}}^{\left( {FD} \right)} + \gamma _{{D_{AF}}}^{\left( {FD} \right)\left( 2 \right)} + 1}},	
	\end{equation}
\end{small}
where ${\gamma _{{D_{AF}}}^{\left( {FD} \right)\left( 2 \right)}}=\gamma_{{D_{DF}}}^{\left( {FD} \right)}$.

\subsection{RIS-Assisted Communication}

Let us denote by ${{\bf{h}}_{S,RIS}} \in \mathbb{C}{^{{N_{RIS}} \times 1}}$ and ${{\bf{h}}_{RIS,D}} \in \mathbb{C}{^{1 \times {N_{RIS}}}}$ the fast-fading complex envelopes of the source-to-RIS and RIS-to-destination links, respectively. In addition, ${\bf{\Phi }}  = R_A {\rm{diag}}\left( {{e^{j{\varphi _1}}}, \ldots ,{e^{j{\varphi _{{N_{RIS}}}}}}} \right)$ denotes the diagonal matrix representing the reflective and phase-shifting values of the RIS elements, and $R_A \in \left( {0,1} \right]$ is the amplitude reflection coefficient, where $R_A=1$ corresponds to the ideal case of lossless reflection. Based on the above, the received signal at the destination consists of the sum of the direct link signal and of the signal that is reflected by the RIS. By denoting it by $y_D^{RIS}$, it holds that
\begin{small}
	\begin{equation}
y_D^{RIS} = \sqrt {{p_{tot}}{P_L}\left( {{d_{SD}}} \right)} {h_{SD}}s + \sqrt {{p_{tot}}{P_L^{RIS}}} {\bf{h}}_{S,RIS}^T{\bf{\Phi h}}_{RIS,D}^Ts + {n_D},
\end{equation}
\end{small}
where $P_L^{RIS}$ is the path-loss in the RIS case.
Hence, the resulting SNR, which we denote by $\gamma_D^{RIS}$, is given by
\begin{small}
	\begin{equation}
\gamma _D^{RIS} = \frac{{{{\left| {\sqrt {{p_{tot}}{P_L}\left( {{d_{SD}}} \right)} {h_{SD}} + \sqrt {{p_{tot}}P_L^{RIS}} {\bf{h}}_{S,RIS}^T{\bf{\Phi h}}_{RIS,D}^T} \right|}^2}}}{{{N_0}}}
\end{equation}
\end{small}
The maximum $\gamma _D^{RIS}$, which we denote by $\gamma _{D_{max}}^{RIS}$, is achieved by adjusting the phase shifts ${\varphi _1}, \ldots ,{\varphi _{{N_{RIS}}}}$ in a way that the direct path signal and the reflected signals from the RIS are co-phased. Based on this, the optimal phase shifts employed by the RIS, which we denote by ${\varphi _1^{opt}}, \ldots ,{\varphi _{{N_{RIS}}}^{opt}}$, are given by
\begin{small}
	\begin{equation} 
\left( {\begin{array}{*{20}{c}}
	{\varphi _1^{opt}}& \cdots &{\varphi _{{N_{RIS}}}^{opt}}
	\end{array}} \right) = \left( {\begin{array}{*{20}{c}}
	{\arg \left[ {{h_{SD}}} \right] - \arg \left[ {{{\left[ {{h_{S,RIS}}} \right]}_1}{{\left[ {{h_{RIS,D}}} \right]}_1}} \right]}& \cdots &{\arg \left[ {{h_{SD}}} \right] - \arg \left[ {{{\left[ {{h_{S,RIS}}} \right]}_{{N_{RIS}}}}{{\left[ {{h_{RIS,D}}} \right]}_{{N_{RIS}}}}} \right]}
	\end{array}} \right).
\end{equation}
\end{small}
As a result, it holds that
\begin{small}
	\begin{equation}
\gamma _{{D_{\max }}}^{RIS} = \frac{{{p_{tot}}{{\left( {\sqrt {{P_L}\left( {{d_{SD}}} \right)} \left| {{h_{SD}}} \right| + \sqrt {P_L^{RIS}} R_A\sum\limits_{n = 1}^{{N_{RIS}}} {{{\left[ {{h_{S,RIS}}} \right]}_n}{{\left[ {{h_{RIS,D}}} \right]}_n}} } \right)}^2}}}{{{N_0}}}.
\end{equation}
\end{small}
As far as $P_L^{RIS}$ and the resulting instantaneous maximum SNR are concerned, we consider the two extreme cases where the elements of the RIS act as anomalous reflectors and diffuse scatterers.

\subsubsection{\textbf{Anomalous Reflection}}

\label{description_of_anomalous_reflection_case}

This case occurs when the size of the RIS elements is sufficiently large {as compared with the wavelength of the radio wave (at least one order of magnitude larger)} so that they act as anomalous reflectors, {according to the generalized Snell's law}, and the theory of geometric optics can be applied to model the interactions between the RIS and the signals \cite{MDR_Access}. {This means that the RIS can reflect the impinging wave towards an arbitrary angle instead of the one predicted by the conventional Snell's law (specular reflection)}. In such a case, according to the theory of geometric optics, the received power under free-space propagation is expected to scale with $\left[\left({d_{SR}}+{d_{RD}}\right)^2\right]^{-1}$, which is the sum-distance law {of a specular reflection \cite[p.~12]{Modeling_The_Wireless_Propagation_Channel}}. Hence, it holds that $P_L^{RIS}\approx{P_L}\left( {d_{SR}}+{d_{RD}} \right)$. As a result, the instantaneous maximum SNR, which we denote by $\gamma _{{D_{\max }}}^{RIS,\; anomalous\; reflection}$, is approximated as
\begin{small}
	\begin{equation}
	\gamma _{{D_{\max }}}^{RIS,\; anomalous\; reflection} \approx \frac{{{p_{tot}}{{\left( {\sqrt {{P_L}\left( {{d_{SD}}} \right)} \left| {{h_{SD}}} \right| + \sqrt {{P_L}\left( {{d_{SR}} + {d_{RD}}} \right)} R_A\sum\limits_{n = 1}^{{N_{RIS}}} {{{\left[ {{h_{S,RIS}}} \right]}_n}{{\left[ {{h_{RIS,D}}} \right]}_n}} } \right)}^2}}}{{{N_0}}}.
	\end{equation}
\end{small}

\subsubsection{\textbf{Diffuse Scattering}}

\label{description_of_diffuse_scattering_case}

This case occurs when the size of the elements of the RIS is comparable {(of the same order of magnitude or smaller)} with the wavelength. In such a case, the elements of the RIS are expected to act as diffusers (dipole scatterers) {\cite[p.~12]{Modeling_The_Wireless_Propagation_Channel}}. {As a result}, the received power under free-space propagation is expected to scale with $\left({d_{SR}^2}{d_{RD}^2}\right)^{-1}$, {which is the product-distance law of backscattering communications, such as radar \cite[p.~12]{Modeling_The_Wireless_Propagation_Channel}}. Hence, it holds that $P_L^{RIS}\approx{P_L}\left( {d_{SR}}\right){P_L}\left({d_{RD}} \right)$. As a result, the instantaneous maximum SNR, which we denote by $\gamma _{{D_{\max }}}^{RIS,\; diffuse\; scattering}$, is approximated as
\begin{small}
	\begin{equation}
\gamma _{{D_{\max }}}^{RIS,\; diffuse\; scattering} \approx \frac{{{p_{tot}}{{\left( {\sqrt {{P_L}\left( {{d_{SD}}} \right)} \left| {{h_{SD}}} \right| + \sqrt {{P_L}\left( {d_{SR}}\right){P_L}\left({d_{RD}} \right)} R_A\sum\limits_{n = 1}^{{N_{RIS}}} {{{\left[ {{h_{S,RIS}}} \right]}_n}{{\left[ {{h_{RIS,D}}} \right]}_n}} } \right)}^2}}}{{{N_0}}}.
 \end{equation}
\end{small}

The end-to-end SNR formulas for the relay-assisted and RIS-assisted communication are summarized in Table~\ref{SNR_expressions}.

\begin{table*}[!t]
	\label{SNR_expressions}
	\caption{End-to-end SNR of relay-assisted and RIS-assisted communication} 
	\centering 
	\scalebox{0.85}{
		\begin{tabular}{| c | c | c | } 
			\hline
			HD relaying under the DF protocol&  
			$\gamma _{{R_{DF}}}^{\left( {HD} \right)} = \frac{{{p_S}{P_L}\left( {{d_{SR}}} \right){{\left\| {{{\bf{h}}_{SR}}} \right\|}^2}}}{{{N_0}}},{\;\;\gamma _{{D_{DF}}}^{\left( {HD} \right)\left( 1 \right)}=\frac{{{p_S}{P_L}\left( {{d_{SD}}} \right){{\left| {{h_{SD}}} \right|}^2}}}{{{N_0}}},\;\; \gamma _{{D_{DF}}}^{\left( {HD} \right)\left( 2 \right)}=\frac{{{p_R}{P_L}\left( {{d_{RD}}} \right){{\left\| {{{\bf{h}}_{RD}}} \right\|}^2}}}{{{N_0}}}}$
			\\ [0.5ex]
			\hline
			HD relaying under the AF protocol&$\gamma _{{D_{AF}}}^{\left( {HD} \right)} = \frac{{{p_S}{P_L}\left( {{d_{SD}}} \right){{\left| {{h_{SD}}} \right|}^2}}}{{{N_0}}}+\frac{{\frac{{{p_S}{P_L}\left( {{d_{SR}}} \right){{\left\| {{{\bf{h}}_{SR}}} \right\|}^2}}}{{{N_0}}}\frac{{{p_R}{P_L}\left( {{d_{RD}}} \right){{\left\| {{{\bf{h}}_{RD}}} \right\|}^2}}}{{{N_0}}}}}{{\frac{{{p_S}{P_L}\left( {{d_{SR}}} \right){{\left\| {{{\bf{h}}_{SR}}} \right\|}^2}}}{{{N_0}}} + \frac{{{p_R}{P_L}\left( {{d_{RD}}} \right){{\left\| {{{\bf{h}}_{RD}}} \right\|}^2}}}{{{N_0}}} + 1}}$ \\ [0.5ex]
			\hline
			FD relaying under the DF protocol&$\gamma _{{R_{DF}}}^{\left( {FD} \right)}=\frac{{\frac{{{p_S}{P_L}\left( {{d_{SR}}} \right){{\left\| {{{\bf{\tilde h}}_{SR}}} \right\|}^2}}}{{{N_0}}}}}{{\frac{{{p_R}{{\left| {{{\bf{w}}_{{R_{MRC}}}^{{\left(FD\right)}}}{{{\bf{\tilde h}}}_{LI}}} \right|}^2}}}{{{N_0}}} + 1}},\;\;\gamma _{{D_{DF}}}^{\left( {FD} \right)}={\frac{{\frac{{{p_R}{P_L}\left( {{d_{RD}}} \right){{\left\| {{{\bf{\tilde h}}_{RD}}} \right\|}^2}}}{{{N_0}}}}}{{\frac{{{p_S}{P_L}\left( {{d_{SD}}} \right){{\left| {{h_{SD}}} \right|}^2}}}{{{N_0}}} + 1}}}$ \\ [0.5ex]
			\hline
			FD relaying under the AF protocol&$\gamma _{{D_{AF}}}^{\left( {FD} \right)} = \frac{{\frac{{\frac{{{p_S}{P_L}\left( {{d_{SR}}} \right){{\left\| {{{\bf{\tilde h}}_{SR}}} \right\|}^2}}}{{{N_0}}}}}{{\frac{{{p_R}{{\left| {{{\bf{w}}_{{R_{MRC}}}^{\left(FD\right)}}{{{\bf{\tilde h}}}_{LI}}} \right|}^2}}}{{{N_0}}} + 1}}\frac{{\frac{{{p_R}{P_L}\left( {{d_{RD}}} \right){{\left\| {{{\bf{\tilde h}}_{RD}}} \right\|}^2}}}{{{N_0}}}}}{{\frac{{{p_S}{P_L}\left( {{d_{SD}}} \right){{\left| {{h_{SD}}} \right|}^2}}}{{{N_0}}} + 1}}}}{{\frac{{\frac{{{p_S}{P_L}\left( {{d_{SR}}} \right){{\left\| {{{\bf{\tilde h}}_{SR}}} \right\|}^2}}}{{{N_0}}}}}{{\frac{{{p_R}{{\left| {{{\bf{w}}_{{R_{MRC}}}^{\left(FD\right)}}{{{\bf{\tilde h}}}_{LI}}} \right|}^2}}}{{{N_0}}} + 1}} + \frac{{\frac{{{p_R}{P_L}\left( {{d_{RD}}} \right){{\left\| {{{\bf{\tilde h}}_{RD}}} \right\|}^2}}}{{{N_0}}}}}{{\frac{{{p_S}{P_L}\left( {{d_{SD}}} \right){{\left| {{h_{SD}}} \right|}^2}}}{{{N_0}}} + 1}} + 1}}$ \\ [0.5ex]
			\hline
			RIS as an anomalous reflector &$\gamma _{{D_{\max }}}^{RIS,\; anomalous\; reflection}\approx\frac{{{p_{tot}}{{\left( {\sqrt {{P_L}\left( {{d_{SD}}} \right)} \left| {{h_{SD}}} \right| + \sqrt {{P_L}\left( {{d_{SR}}+{d_{RD}}} \right)} R_A\sum\limits_{n = 1}^{{N_{RIS}}} {{{\left[ {{h_{S,RIS}}} \right]}_n}{{\left[ {{h_{RIS,D}}} \right]}_n}} } \right)}^2}}}{{{N_0}}}$\\ [0.5ex]
			\hline
			RIS as a dipole scatterer &$\gamma _{{D_{\max }}}^{RIS,\; diffuse\; scattering}\approx\frac{{{p_{tot}}{{\left( {\sqrt {{P_L}\left( {{d_{SD}}} \right)} \left| {{h_{SD}}} \right| + \sqrt {{P_L}\left( {d_{SR}}\right){P_L}\left({d_{RD}} \right)} R_A\sum\limits_{n = 1}^{{N_{RIS}}} {{{\left[ {{h_{S,RIS}}} \right]}_n}{{\left[ {{h_{RIS,D}}} \right]}_n}} } \right)}^2}}}{{{N_0}}}$ \\ [0.5ex]
			\hline
	\end{tabular}}
	\label{SNR_expressions} 
\end{table*}

\section{Achievable Rate}

\label{achievable_rate}

In this section, we provide the mathematical formulation for the achievable rate for both relay-assisted and RIS-assisted communication. For the relaying case, we derive the optimal allocation of $p_S$ and $p_R$ so that the achievable rate is maximized.

\subsection{Relay-Assisted Communication}

Let us first define the following parameters that are used in the computation of the optimal power allocation for each of the following case studies.

\begin{small}
	\begin{equation}
	\begin{gathered}
	\begin{aligned}
    A &= \frac{{{P_L}\left( {{d_{SD}}} \right){{\left| {{h_{SD}}} \right|}^2}}}{{{N_0}}}\\
    {B^{\left( {HD} \right)}} &= \frac{{{P_L}\left( {{d_{SR}}} \right){{\left\| {{{\bf{h}}_{SR}}} \right\|}^2}}}{{{N_0}}}\\
    {C^{\left( {HD} \right)}} &= \frac{{{P_L}\left( {{d_{RD}}} \right){{\left\| {{{\bf{h}}_{RD}}} \right\|}^2}}}{{{N_0}}}\\
    {B^{\left( {FD} \right)}} &= \frac{{{P_L}\left( {{d_{SR}}} \right){{\left\| {{{\bf{\tilde h}}_{SR}}} \right\|}^2}}}{{{N_0}}}\\
    {C^{\left( {FD} \right)}} &= \frac{{{P_L}\left( {{d_{RD}}} \right){{\left\| {{{\bf{\tilde h}}_{RD}}} \right\|}^2}}}{{{N_0}}}\\
    D &= \frac{{{{\left| {{{\bf{w}}_{{R_{MRC}}}^{\left( {FD} \right)}}{{{\bf{\tilde h}}}_{LI}}} \right|}^2}}}{{{N_0}}}.
    \end{aligned}
    \end{gathered}
    \end{equation}
\end{small}

\subsubsection{\textbf{HD Relaying}}

\paragraph{DF Operation} The achievable rate under HD-DF relaying, which we denote by $R_{DF}^{\left( {HD} \right)}$, is given by
\begin{small}
	\begin{equation}
R_{DF}^{\left( {HD} \right)} =\frac{1}{2} {\log _2}\left( {1 + \min \left( {\gamma _{{R_{DF}}}^{\left( {HD} \right)},\gamma _{{D_{DF}}}^{\left( {HD} \right)\left( 1 \right)} + \gamma _{{D_{DF}}}^{\left( {HD} \right)\left( 2 \right)}} \right)} \right).
\end{equation}
\end{small}
The maximum rate, which we denote by $R_{DF_{max}}^{\left( {HD} \right)}$, is given by
\begin{small}
	\begin{equation}
	\label{max_rate_HD_DF_Relaying}
R_{D{F_{\max }}}^{\left( {HD} \right)} = \mathop {\max }\limits_{{p_S},{p_R}} {\frac{1}{2}\log _2}\left( {1 + \min \left( {\gamma _{{R_{DF}}}^{\left( {HD} \right)},\gamma _{{D_{DF}}}^{\left( {HD} \right)\left( 1 \right)} + \gamma _{{D_{DF}}}^{\left( {HD} \right)\left( 2 \right)}} \right)} \right)={\frac{1}{2}\log _2}\left( {1 + \mathop {\max }\limits_{{p_S},{p_R}}\min \left( {\gamma _{{R_{DF}}}^{\left( {HD} \right)},\gamma _{{D_{DF}}}^{\left( {HD} \right)\left( 1 \right)} + \gamma _{{D_{DF}}}^{\left( {HD} \right)\left( 2 \right)}} \right)} \right),
\end{equation}
\end{small}
where the last equality holds due to the concavity of the log function. We distinguish the following two cases:

\textbf{Case 1}: $\gamma _{{D_{DF}}}^{\left( {HD} \right)\left( 1 \right)} > \gamma _{{R_{DF}}}^{\left( {HD} \right)}$. In this case, relaying is suboptimal since the direct source-to-destination link achieves a higher rate. Consequently, $p_S=p_{tot}$ and the achievable rate $R_{SISO}$ is given by
\begin{small}
	\begin{equation}
{R_{SISO}} = {\log _2}\left( {1 + \gamma _{{D_{DF}}}^{\left( {HD} \right)\left( 1 \right)}} \right).
\end{equation}
\end{small}

\textbf{Case 2}: $\gamma _{{D_{DF}}}^{\left( {HD} \right)\left( 1 \right)} \le \gamma _{{R_{DF}}}^{\left( {HD} \right)}$. In this case, the term $\min \left( {\gamma _{{R_{DF}}}^{\left( {HD} \right)},\gamma _{{D_{DF}}}^{\left( {HD} \right)\left( 1 \right)} + \gamma _{{D_{DF}}}^{\left( {HD} \right)\left( 2 \right)}} \right)$ in 
\eqref{max_rate_HD_DF_Relaying} is maximized for $\gamma _{{R_{DF}}}^{\left( {HD} \right)}=\gamma _{{D_{DF}}}^{\left( {HD} \right)\left( 1 \right)} + \gamma _{{D_{DF}}}^{\left( {HD} \right)\left( 2 \right)}$. The obtained optimal value of $p_S$, which we denote by $p_{S_{DF}}^{opt(HD)}$, is given by
\begin{small}
	\begin{equation}
p_{S_{DF}}^{opt\left( {HD} \right)} = \frac{{{p_{tot}}{C^{\left( {HD} \right)}}}}{{{B^{\left( {HD} \right)}} - A + {C^{\left( {HD} \right)}}}}.
\end{equation}
\end{small}
The optimal value of $p_R$, which we denote by $p_{R_{DF}}^{opt\left( {HD} \right)}$, is equal to $p_{R_{DF}}^{opt\left( {HD} \right)}=p_{tot}-p_{S_{DF}}^{opt\left( {HD} \right)}$.

\paragraph{AF Operation} The achievable rate under HD-AF relaying, which we denote by $R_{AF}^{\left( {HD} \right)}$, is given by
\begin{small}
	\begin{equation}
	R_{AF}^{\left( {HD} \right)} = \frac{1}{2}{\log _2}\left( {1 + \gamma _{{D_{AF}}}^{\left( {HD} \right)}} \right).
    \end{equation}
\end{small}
The maximum rate, which we denote by $R_{AF_{max}}^{\left( {HD} \right)}$, is given by
\begin{small}
	\begin{equation}
	R_{A{F_{\max }}}^{\left( {HD} \right)} = \mathop {\max }\limits_{{p_S},{p_R}} \frac{1}{2}{\log _2}\left( {1 + \gamma _{{D_{AF}}}^{\left( {HD} \right)}} \right) = \frac{1}{2}{\log _2}\left( {1 + \mathop {\max }\limits_{{p_S},{p_R}} \gamma _{{D_{AF}}}^{\left( {HD} \right)}} \right).
	\end{equation}
\end{small}
After replacing $p_R$ at $\gamma _{{D_{AF}}}^{\left( {HD} \right)}$ with $p_{tot}-p_S$ and taking the first-order derivative with respect to $p_S$ and setting it to zero, the optimum value of $p_S$, which we denote by $p_{S_{AF}}^{opt\left( {HD} \right)}$, is given by
\begin{tiny}
	\begin{equation}
	p_{{S_{AF}}}^{opt\left( {HD} \right)} = \frac{{\left[ {AB^{\left(HD\right)} - \left( {A + B^{\left(HD\right)}} \right)C^{\left(HD\right)}} \right]\left( {1 + C^{\left(HD\right)}{p_{tot}}} \right) \pm \sqrt {B^{\left(HD\right)}C^{\left(HD\right)}\left[ {\left( {A + B^{\left(HD\right)}} \right)C^{\left(HD\right)} - AB^{\left(HD\right)}} \right]\left( {1 + B^{\left(HD\right)}{p_{tot}}} \right)\left( {1 + C^{\left(HD\right)}{p_{tot}}} \right)} }}{{\left( {B^{\left(HD\right)} - C^{\left(HD\right)}} \right)\left[ {\left( {A + B^{\left(HD\right)}} \right)C^{\left(HD\right)} - AB^{\left(HD\right)}} \right]}}.
	\end{equation}
\end{tiny}  
From the two values of $p_{{S_{AF}}}^{opt\left( {HD} \right)}$, we keep the one for which it holds that $0 < p_{{S_{AF}}}^{opt\left( {HD} \right)} \le {p_{tot}}$. The optimal value of $p_R$, which we denote by $p_{R_{AF}}^{opt\left( {HD} \right)}$, is equal to $p_{R_{AF}}^{opt\left( {HD} \right)}=p_{tot}-p_{S_{AF}}^{opt\left( {HD} \right)}$.

\subsubsection{\textbf{FD Relaying}}

\paragraph{DF Operation}
The achievable rate under FD-DF relaying, which we denote by $R_{DF}^{\left( {FD} \right)}$, is given by
\begin{small}
	\begin{equation}
R_{DF}^{\left( {FD} \right)} = {\log _2}\left( {1 + \min \left( {\gamma _{{R_{DF}}}^{\left( {FD} \right)},\gamma _{{D_{DF}}}^{\left( {FD} \right)}} \right)} \right).
\end{equation}
\end{small}
The maximum achievable rate, which we denote by $R_{DF_{max}}^{\left( {FD} \right)}$, is given by
\begin{small}
	\begin{equation}
	R_{DF_{max}}^{\left( {FD} \right)}=\mathop {\max }\limits_{{p_S},{p_R}}{\log _2}\left( {1 + \min \left( {\gamma _{{R_{DF}}}^{\left( {FD} \right)},\gamma _{{D_{DF}}}^{\left( {FD} \right)}} \right)} \right)={\log _2}\left( {1 + \mathop {\max }\limits_{{p_S},{p_R}}\min \left( {\gamma _{{R_{DF}}}^{\left( {FD} \right)},\gamma _{{D_{DF}}}^{\left( {FD} \right)}} \right)} \right).
	\end{equation}
\end{small}
Hence, $R_{DF}^{\left( {FD} \right)}$ is maximized if $p_S$ and $p_R$ take the values for which ${\gamma _{{R_{DF}}}^{\left( {FD} \right)}=\gamma _{{D_{DF}}}^{\left( {FD} \right)}}$ holds. Accordingly, the optimum value of $p_S$, which we denote by $p_{S_{DF}}^{opt\left( {FD} \right)}$, is given by
\begin{small}
	\begin{equation}
p_{{S_{DF}}}^{opt\left( {FD} \right)} = \frac{{ - b_{DF}^{\left( {FD} \right)} \pm \sqrt {{{\left( {b_{DF}^{\left( {FD} \right)}} \right)}^2} - 4a_{DF}^{\left( {FD} \right)}c_{DF}^{\left( {FD} \right)}} }}{{2a_{DF}^{\left( {FD} \right)}}},
	\end{equation}
\end{small} 
where 
\begin{small}
	\begin{equation}
	\begin{gathered}
	\begin{aligned}
a_{DF}^{\left( {FD} \right)}&=A{B^{\left( {FD} \right)}} - D{C^{\left( {FD} \right)}}\\
b_{DF}^{\left( {FD} \right)} &= {B^{\left( {FD} \right)}} + {C^{\left( {FD} \right)}} + 2{p_{tot}}D{C^{\left( {FD} \right)}}.\\
c_{DF}^{\left( {FD} \right)} &=  - p_{tot}^2D{C^{\left( {FD} \right)}} - {p_{tot}}{C^{\left( {FD} \right)}}
\end{aligned} 
\end{gathered}
\end{equation}
\end{small}
From the two values of $p_{{S_{DF}}}^{opt\left( {FD} \right)}$, we keep the one for which it holds that $0 < p_{{S_{DF}}}^{opt\left( {FD} \right)} \le {p_{tot}}$. The optimal value of $p_R$, which we denote by $p_{R_{DF}}^{opt\left( {FD} \right)}$, is equal to $p_{R_{DF}}^{opt\left( {FD} \right)}=p_{tot}-p_{S_{DF}}^{opt\left( {FD} \right)}$.

\paragraph{AF Operation}
The achievable rate under FD-AF relaying, which we denote by $R_{AF}^{\left( {FD} \right)}$, is given by
\begin{small}
	\begin{equation}
	R_{AF}^{\left( {FD} \right)} = {\log _2}\left( {1 + \gamma _{{D_{AF}}}^{\left( {FD} \right)}} \right).
	\end{equation}
	\end{small}
The maximum achievable rate, which we denote by $R_{DF_{max}}^{\left( {FD} \right)}$, is given by
\begin{small}
	\begin{equation}
	R_{AF_{max}}^{\left( {FD} \right)}=\mathop {\max }\limits_{{p_S},{p_R}}{\log _2}\left( {1 + \gamma _{{D_{AF}}}^{\left( {FD} \right)}} \right)={\log _2}\left( {1 + \mathop {\max }\limits_{{p_S},{p_R}}\gamma _{{D_{AF}}}^{\left( {FD} \right)}} \right).
	\end{equation}
\end{small}
Consequently, to maximize $R_{AF_{max}}^{\left( {FD} \right)}$ we replace $p_R$ at $\gamma _{{D_{AF}}}^{\left( {FD} \right)}$ with $p_{tot}-p_S$ and then we take the first-order derivative with respect to $p_S$ and set it to zero. According to the solution, the optimum value of $p_S$, which we denote by $p_{S_{AF}}^{opt\left( {FD} \right)}$, is given by
\begin{small}
	\begin{equation}
	p_{{S_{AF}}}^{opt\left( {FD} \right)} = \frac{{ - b_{AF}^{\left( {FD} \right)} \pm \sqrt {{{\left( {b_{AF}^{\left( {FD} \right)}} \right)}^2} - 4a_{AF}^{\left( {FD} \right)}c_{AF}^{\left( {FD} \right)}} }}{{2a_{AF}^{\left( {FD} \right)}}},
	\end{equation}
\end{small} 
where 
\begin{small}
	\begin{equation}
	\begin{gathered}
	\begin{aligned}
	a_{AF}^{\left( {FD} \right)}&=\left( {A{B^{\left( {FD} \right)}} - {C^{\left( {FD} \right)}}D} \right){p_{tot}} + A + {B^{\left( {FD} \right)}} - {C^{\left( {FD} \right)}} - D\\
	b_{AF}^{\left( {FD} \right)} &= 2{C^{\left( {FD} \right)}}Dp_{tot}^2 + 2\left( {{C^{\left( {FD} \right)}} + D} \right){p_{tot}} + 2.\\
	c_{AF}^{\left( {FD} \right)} &= - {C^{\left( {FD} \right)}}Dp_{tot}^3 - \left( {{C^{\left( {FD} \right)}} + D} \right)p_{tot}^2 - {p_{tot}}
	\end{aligned} 
	\end{gathered}
	\end{equation}
\end{small}
From the two values of $p_{{S_{AF}}}^{opt\left( {FD} \right)}$, we keep the one for which it holds that $0 < p_{{S_{AF}}}^{opt\left( {FD} \right)} \le {p_{tot}}$. The optimal value of $p_R$, which we denote by $p_{R_{AF}}^{opt\left( {FD} \right)}$, is equal to $p_{R_{AF}}^{opt\left( {FD} \right)}=p_{tot}-p_{S_{AF}}^{opt\left( {FD} \right)}$.

\subsection{RIS-Assisted Communication}

As mentioned, RISs are an emerging field of research in wireless communications. The interested readers can refer to \cite{MDR_Eurasip} and \cite{MDR_Access} for further information. Depending on the ratio between the geometric size and the wavelength of the radio wave, the following two limiting cases can be considered:

\subsubsection{\textbf{Anomalous Reflection}}
{As mentioned in Section~\ref{description_of_anomalous_reflection_case}}, if the geometric size of the RIS is sufficiently large as compared with the wavelength of the radio wave {(at least one order of magnitude larger)}, then each element of the RIS behaves like a mirror that can reflect the signals towards directions that are different from the direction of arrival. This case study is referred to as anomalous reflection.

The maximum achievable rate in this case, which we denote by $R_{{{\max }}}^{RIS,\; anomalous\; reflection}$, is given by
\begin{small}
	\begin{align}
			R _{{{\max }}}^{RIS,\; anomalous\; reflection}&={\log _2}\left( {1 + \gamma _{{D_{\max }}}^{RIS,\; anomalous\; reflection}} \right)\nonumber\\
			&\approx{\log _2}\left( {1 +  \frac{{{p_{tot}}{{\left( {\sqrt {{P_L}\left( {{d_{SD}}} \right)} \left| {{h_{SD}}} \right| + \sqrt {{P_L}\left( {{d_{SR}}+{d_{RD}}} \right)} R_A\sum\limits_{n = 1}^{{N_{RIS}}} {{{\left[ {{h_{S,RIS}}} \right]}_n}{{\left[ {{h_{RIS,D}}} \right]}_n}} } \right)}^2}}}{{{N_0}}}} \right).
	\end{align}
\end{small}

\subsubsection{\textbf{Diffuse Scattering}}
{As mentioned in Section~\ref{description_of_diffuse_scattering_case}}, if the geometric size of the RIS is of the same order as or smaller than the wavelength of the radio wave, then the RIS behaves as a dipole scatterer that diffuses the signals towards all possible directions. This case study is referred to as diffuse scattering.

The maximum achievable rate in this case, which we denote by $R_{{{\max }}}^{RIS,\; diffuse\; scattering}$, is given by
\begin{small}
	\begin{align}
	R _{{{\max }}}^{RIS,\; diffuse\; scattering}&={\log _2}\left( {1 + \gamma _{{D_{\max }}}^{RIS,\; diffuse\; scattering}} \right)\nonumber\\
	&\approx{\log _2}\left( {1 +  \frac{{{p_{tot}}{{\left( {\sqrt {{P_L}\left( {{d_{SD}}} \right)} \left| {{h_{SD}}} \right| + \sqrt {{P_L}\left( {d_{SR}}\right){P_L}\left({d_{RD}} \right)} R_A\sum\limits_{n = 1}^{{N_{RIS}}} {{{\left[ {{h_{S,RIS}}} \right]}_n}{{\left[ {{h_{RIS,D}}} \right]}_n}} } \right)}^2}}}{{{N_0}}}} \right).
	\end{align}
\end{small}

The formulas of the achievable rate for the relay-assisted and RIS-assisted communication are summarized in Table~\ref{rate_expressions}.

\begin{table*}[!t]
	\label{rate_expressions}
	\caption{Achievable rate of the relay-assisted and RIS-assisted communication} 
	\centering 
	\scalebox{0.80}{
		\begin{tabular}{| c | c | c | } 
			\hline
			HD relaying under the DF protocol& 
				$R_{D{F_{\max }}}^{\left( {HD} \right)} ={\frac{1}{2}\log _2}\left( {1 + \mathop {\max }\limits_{{p_S},{p_R}}\min \left( {\gamma _{{R_{DF}}}^{\left( {HD} \right)},\gamma _{{D_{DF}}}^{\left( {HD} \right)\left( 1 \right)} + \gamma _{{D_{DF}}}^{\left( {HD} \right)\left( 2 \right)}} \right)} \right)$, if $\gamma _{{D_{DF}}}^{\left( {HD} \right)\left( 1 \right)} \le \gamma _{{R_{DF}}}^{\left( {HD} \right)}$ \\ 
			\hline
			HD relaying under the AF protocol&$R_{A{F_{\max }}}^{\left( {HD} \right)} = \frac{1}{2}{\log _2}\left( {1 + \mathop {\max }\limits_{{p_S},{p_R}} \gamma _{{D_{AF}}}^{\left( {HD} \right)}} \right)$ \\ 
			\hline
			FD relaying under the DF protocol&$R_{DF_{max}}^{\left( {FD} \right)}={\log _2}\left( {1 + \mathop {\max }\limits_{{p_S},{p_R}}\min \left( {\gamma _{{R_{DF}}}^{\left( {FD} \right)},\gamma _{{D_{DF}}}^{\left( {FD} \right)}} \right)} \right)$ \\ 
			\hline
			FD relaying under the AF protocol&$R_{AF_{max}}^{\left( {FD} \right)}={\log _2}\left( {1 + \mathop {\max }\limits_{{p_S},{p_R}}\gamma _{{D_{AF}}}^{\left( {FD} \right)}} \right)$ \\ 
			\hline
			RIS as an anomalous reflector &$R _{{{\max }}}^{RIS,\; anomalous\; reflection}\approx{\log _2}\left( {1 +  \frac{{{p_{tot}}{{\left( {\sqrt {{P_L}\left( {{d_{SD}}} \right)} \left| {{h_{SD}}} \right| + \sqrt {{P_L}\left( {{d_{SR}}+{d_{RD}}} \right)} R_A\sum\limits_{n = 1}^{{N_{RIS}}} {{{\left[ {{h_{S,RIS}}} \right]}_n}{{\left[ {{h_{RIS,D}}} \right]}_n}} } \right)}^2}}}{{{N_0}}}} \right)$\\ 
			\hline
			RIS as a dipole scatterer &$R _{{{\max }}}^{RIS,\; diffuse\; scattering}\approx{\log _2}\left( {1 +  \frac{{{p_{tot}}{{\left( {\sqrt {{P_L}\left( {{d_{SD}}} \right)} \left| {{h_{SD}}} \right| + \sqrt {{P_L}\left( {d_{SR}}\right){P_L}\left({d_{RD}} \right)} R_A\sum\limits_{n = 1}^{{N_{RIS}}} {{{\left[ {{h_{S,RIS}}} \right]}_n}{{\left[ {{h_{RIS,D}}} \right]}_n}} } \right)}^2}}}{{{N_0}}}} \right)$ \\ 
			\hline
	\end{tabular}}
	\label{rate_expressions} 
\end{table*}

\section{Conclusion}
In this report, we have summarized the end-to-end signal-to-noise ratio and achievable rate formulas of half-duplex, full-duplex, amplify-and-forward, and decode-and-forward relay-assisted communication, as well as the signal-to-noise ratio and achievable rate formulas of the emerging technology known as reconfigurable intelligent surfaces. The obtained formulas constitute the departing point for comparing wireless communications assisted by relays and reconfigurable intelligent surfaces.


\bibliographystyle{IEEEtran}

\end{document}